\documentclass[letterpaper]{article} 
\usepackage{aaai25}  
\usepackage{times}  
\usepackage{helvet}  
\usepackage{courier}  
\usepackage[hyphens]{url}  
\usepackage{graphicx} 
\usepackage{natbib}  
\usepackage{caption} 
\usepackage{algorithm}
\usepackage{algorithmic}

\usepackage{tabularx}
\usepackage{amsmath}
\usepackage{multirow}
\usepackage{newfloat}
\usepackage{listings}
\usepackage{xcolor}
\DeclareCaptionStyle{ruled}{labelfont=normalfont,labelsep=colon,strut=off} 
\floatstyle{ruled}
\newfloat{listing}{tb}{lst}{}
\floatname{listing}{Listing}
\usepackage{booktabs}  
\usepackage{caption}   

\title{Informing AI Risk Assessment with News Media: Analyzing National and Political Variation in the Coverage of AI Risks}
\author {
    Mowafak Allaham\textsuperscript{\rm 1},
    Kimon Kieslich \textsuperscript{\rm 2},
    Nicholas Diakopoulos\textsuperscript{\rm 1}
}
\affiliations {
    \textsuperscript{\rm 1}Northwestern University\\
    \textsuperscript{\rm 2}Institute for Information Law, University of Amsterdam \\
    mowafakallaham2021@u.northwestern.edu, k.kieslich@uva.nl, nad@northwestern.edu
}

\usepackage{bibentry}

\begin{document}

\maketitle

\begin{abstract}
Risk-based approaches to AI governance often center the technological artifact as the primary focus of risk assessments, overlooking systemic risks that emerge from the complex interaction between AI systems and society. One potential source to incorporate more societal context into these approaches is the news media, as it embeds and reflects complex interactions between AI systems, human stakeholders, and the larger society. News media is influential in terms of which AI risks are emphasized and discussed in the public sphere, and thus which risks are deemed important. Yet, variations in the news media between countries and across different value systems (e.g. political orientations) may differentially shape the prioritization of risks through the media's agenda setting and framing processes. To better understand these variations, this work presents a comparative analysis of a cross-national sample of news media spanning 6 countries (the U.S., the U.K., India, Australia, Israel, and South Africa). Our findings show that AI risks are prioritized differently across nations and shed light on how left vs. right leaning U.S. based outlets not only differ in the prioritization of AI risks in their coverage, but also use politicized language in the reporting of these risks. These findings can inform risk assessors and policy-makers about the nuances they should account for when considering news media as a supplementary source for risk-based governance approaches.
\end{abstract}

%

\section{Introduction}

AI is increasingly integrated into various systems and applications that serve millions of users globally, despite its potential for significant risks to both individuals and society \cite{bommasani2021opportunities,park2024ai,burtell2023artificial}. In response, governments, companies, and researchers have contributed regulatory frameworks, \cite{madiega2021artificial,biden2023executive}, risk assessments \cite{solaiman_evaluating_2023, metcalf_algorithmic_2021,allaham2024towards,nanayakkara_unpacking_2021} and assessment methods, such as red-teaming and safety benchmarks \cite{ganguli2022red,mazeika2024harmbench, zeng_ai_2024,zhang2023safetybench}, to help govern \cite{reuel2024open}, anticipate \cite{kieslich_anticipating_2024,hautala_spectrum_2023,avin_exploring_2020}, and potentially mitigate such risks. Risk-based approaches \cite{van_der_heijden_risk_2021} to managing potential harms from AI have come to dominate, yet such approaches often center the technological artifact as the primary focus of risk assessments, overlooking systemic risks of AI that see risk arising as a complex interaction between the AI system, human stakeholders (e.g. users but also impacted people), and the larger societal context which might include dimensions of culture, government, institutions, and more \cite{kieslich2025scenario, weidinger2023sociotechnical, uuk_taxonomy_2024}. 

One route to incorporate more societal context into risk-based approaches is to leverage the news media. As observed in AI incident tracking initiatives such as the AI Incident Database and OECD AI Incidents Monitor \cite{mcgregor2021preventing,filippucci2024impact,diakopoulos_2025}, news coverage, including the reporting of failures, biases, and broader impacts of algorithmic and AI systems on the public \cite{diakopoulos2015algorithmic} can be leveraged as a source to help map AI risks in real-world contexts. In particular news media functions to identify and articulate risks and harms in the full complexity of society, emphasizing the socio-technical interactions around AI systems \cite{diakopoulos_2025}. At the same time, news media and the journalistic processes underlying it, reflects its own set of normative and other biases about what is prioritized for coverage and how AI is covered \cite{chuan2019framing,nguyen_news_2024}. Moreover, news media plays an important role in shaping the narrative around AI risks that are relevant to the general public and the perception of AI by various stakeholders including the public and policy-makers \cite{gilardi2024we}. Thus, differences in the news media between societies (e.g. as indicated by country) and across different value systems (e.g. political orientations) may differentially shape the identification and perception of harms through agenda setting and framing processes \cite{ouchchy2020ai,sun2020newspaper,brennen2018industry, mccombs1972agenda,scheufele1999framing}.

Owing to the role of news media in shaping how risks are perceived by society, in this work we posit that risk assessors, policy-makers, and third-party auditors who evaluate and address the negative impacts of AI through risk-based and regulatory means should be aware of and account for media variance, including national and political nuances, as they consider news media in risk-based regulatory approaches. To shed light on how these details may influence the reporting of AI risks, this work analyzes a sample of news media to examine the prevalence of AI risks, making cross-national comparisons of news reporting between countries from the Global North and Global South, and analyzing the role of political orientation of news outlets in shaping the coverage of AI risks in the U.S. context specifically. 

Using the domain taxonomy of AI risks from the the MIT Risk Repository \cite{slattery2024ai} -- a repository of AI risks synthesizing 56 taxonomies that categorizes AI risks by their cause and risk domain -- we analyze the prevalence of AI risks reported in a sample of news articles published in English by news outlets spanning 6 countries from around the world: the United States of America, the United Kingdom, India, Australia, Israel, and South Africa. Furthermore, we examine how the prevalence and coverage of AI risks varies across U.S. news outlets with different political orientations, using domain-level ratings from Media Bias Fact Check, an independent website maintained by researchers and journalists that relies on human fact-checkers affiliated with the International Fact-Checking Network to evaluate media sources along different dimensions including political bias \cite{lin2023high}. 

Through a comparative analysis of the prevalence of AI risks reported in our sample, we show how news media tend to emphasize the coverage of specific risks such as \textit{Socioeconomic \& Environmental} risks overall, but place less attention on other risks such as \textit{Human-Computer Interaction} risks. We also illustrate the heterogeneous pattern of news coverage for AI risks showing notable and significant associations between countries and AI risks reported in articles published by news outlets in these countries. For instance, even among the most prevalent risk category in our sample, the proportion of articles covering \textit{Socioeconomic \& Environmental} risks is substantially less in Israeli and Indian outlets as compared to those in the United States, the United Kingdom, South Africa, and Australia. Furthermore, we illustrate how the prioritization and communication of AI risks are influenced and shaped by the political orientation of news outlets in our sample of articles from the U.S. Specifically, we show how \textit{Malicious Actors \& Misuse} risks are the most salient risks reported on by right-biased news outlets compared to \textit{Socioeconomic \& Environmental} risks by left-biased outlets. 
In addition, as part of this analysis, we share examples illustrating the use of politicized language in the reporting of AI risks by these outlets. 

As news media continue to play a crucial role in highlighting risks and harms relevant to the general public--a key stakeholder in shaping the current and future development of AI regulations and public policy, especially in democratic countries-- and shaping their perception of AI, this work explicitly focuses on examining the influence of national and political variations embedded in journalistic practices on the reporting of AI risks in news media. By accounting for such variations, we unravel insights about which of the AI risks identified by the MIT domain taxonomy of AI risks are dominating the public discourse (and which are not), per our cross-national comparison of the prevalence of AI risks in news media. In addition, by leveraging our sample of articles from the U.S., we illustrate the association between political orientations of news outlets and the risks covered by these outlets. 

Our findings contribute to informing risk assessors, policymakers, and researchers about (1) dimensions that should be accounted for when considering news media as part of risk-assessment, incidents monitoring, and regulatory practices, and (2) the emerging politicized language around AI risks that may influence the public perception of AI and hinder progress on current and future development of AI regulations and policies aiming to make AI systems and technologies more inclusive and safe.

\section{Related Work on Media Coverage of AI}
The news media play an influential role in shaping the national and public discourse on AI by helping to set the standards and expectations for AI accountability \cite{diakopoulos_2025}. In the traditional understanding of communication science, the news media function as agenda setters \cite{mccombs1972agenda}. A key task of the news media is to inform a broad public about politically and socially relevant issues \cite{jamieson_how_2017} and thereby ensure a plurality of voices, i.e. the inclusion of various societal stakeholders. \emph{How} the media portray these technologies is consequential, as media coverage has been shown to influence public opinion \cite{nisbet_knowledge_2002, scheufele_public_2005}, especially for novel technologies like AI \cite{jamieson_recap_2017} -- and public opinion plays a crucial role in technology adoption. On the one hand, citizens act as consumers of AI technology, and the media's portrayal of AI can influence whether or not people are willing to use the technology \cite{ouchchy_ai_2020}. On the other hand, citizens can act as voters and thus influence regulatory aspects \cite{ouchchy_ai_2020, kieslich_ever_2023, kieslich_role_2024}.

One of the main principles of journalistic news quality is to represent a plurality of voices that are relevant to the discourse, and the inclusion of these voices (e.g., activists, academics, civilians, NGOs) can enrich the discourse on AI \cite{brennen_industry-led_2018}. In particular, when it comes to reporting on AI risks, the efforts of investigative journalists have helped shed light on pressing issues such as the child benefit scandal in the Netherlands \cite{constantaras_inside_2023} or the COMPAS recidivism algorithm in the US \cite{angwin_machine_2022}. Indeed, scholars first articulated the idea of ``algorithmic accountability'' as stemming from investigations published in the news media \cite{diakopoulos_2015}, and more recently have argued for the inclusion of journalists in third-party audits of AI systems, as they ``were responsible for uncovering deeply-rooted socio-technical harms in algorithmic systems related mainly to representational harms due to discriminatory design choices.'' \cite{hartmann_addressing_2024}. 
This is supported by the fact that many of the sources of the AI Incident Database \cite{mcgregor2021preventing} are newspaper articles. As a result, news coverage of AI risks plays a key role in exposing the risks of AI systems. Unlike self-reporting by companies (including their research teams), journalists are structurally independent and can uncover novel impacts that may conflict with corporate goals. Their inclusion ``ensure[s] social accountability through domain knowledge and special access to affected communities'' \cite{hartmann_addressing_2024}.

Another important factor when considering the impact of discourse on public perception is its politicization. Scholarly research in this area states that politicization of an issue requires three conditions \cite{de2011no, schattschneider1957intensity}: (1) polarization of the issue, i.e., whether and how prevalent different (political) positions are on the issue. This could also be achieved by different framing or agenda setting of topics related to the issue (e.g. different prevalence of AI risks, or different positions on individual AI issues). (2) The intensity of media coverage. This refers to the visibility of an issue. The more it is covered, the more relevance is attributed to the issue. And (3) The resonance of the issue, i.e. how relevant the issue is in the eyes of the public. Media coverage plays a key role in this regard, as it provides an important arena in which AI is discussed. Several studies of media coverage have found a sharp increase in news coverage of AI in recent years \cite{fast_long-term_2017, vergeer_artificial_2020, ouchchy_ai_2020, ittefaq_global_2025, chuan2019framing}, which satisfies the condition of intensity of coverage. Several scholars have also analyzed the influence of the political leaning of the news outlet on the framing of AI -- with mixed results \cite{brennen_industry-led_2018, roe_what_2023, vergeer_artificial_2020}. For example, for the UK case, \citet{brennen_industry-led_2018} state that: ``Right-leaning outlets highlight issues of economics and geopolitics, including automation, national security, and investment", whereas ``Left-leaning outlets highlight issues of ethics of AI, including discrimination, algorithmic bias, and privacy." However, when Roe and Perking looked at headlines about ChatGPT and AI in general a few years later, they didn't find any evidence of strong polarization \cite{roe_what_2023}. For the Dutch case, \citet{vergeer_artificial_2020} reported that business newspapers were generally more favorable to AI than national newspapers. However, the politicization of the AI \emph{risk debate} in particular has not been explored. Analyzing the politicization of AI risks in terms of political positions is important because it reflects political strategies in terms of regulation or policy enforcement. Furthermore, it shows how citizens who consume politically biased news are informed and perceive the AI risk discourse.

Recognizing the importance of the news media in relation to AI, a significant number of scholars have focused on analyzing how the news discusses AI \cite{brennen_industry-led_2018, chuan2019framing, fast_long-term_2017, ouchchy_ai_2020, kieslich_everything_2022, sun_newspaper_2020, vergeer_artificial_2020, zeng_ai_2024, brennen_what_2022, meisner_risks_2024, roe_what_2023, ittefaq_global_2025, nguyen_new_2022, nguyen_news_2024, bunz_ai_2022}. However, most studies of media coverage focus on countries in the Global North, such as the US \cite{chuan2019framing, fast_long-term_2017}, the UK \cite{brennen_industry-led_2018, brennen_what_2022, roe_what_2023}, Germany \cite{meisner_risks_2024, kieslich_everything_2022}, the Netherlands \cite{vergeer_artificial_2020}, or take a comparative approach between the US and the UK \cite{nguyen_news_2024, bunz_ai_2022} or the US and China \cite{nguyen_new_2022}. Only a few studies focus on non-Western countries, with the exception of China \cite{zeng_ai_2024} and a comparative study of 12 countries, including countries of the Global North and the Global South \cite{ittefaq_global_2025}. Thematically, studies on media coverage mostly focus on mapping the general discourse on AI, for example by analyzing the thematic structure or sentiment of media discourse (e.g. \citet{ittefaq_global_2025,chuan2019framing,brennen_industry-led_2018}), while rarely focusing explicitly on risks or negative impacts (with the exception of \cite{ouchchy_ai_2020, nguyen_news_2024,chuan2019framing}).

Overall, we find that the extant literature on news analysis of AI doesn't tend to engage extensively with AI \textit{risks} and that there is an opportunity to more fully leverage news sources in AI risk assessment practices by incorporating this focus on risk and its intersection with national and political variations. This paper addresses these opportunities by (1) explicitly focusing on AI risks in the study of news content, (2) using news media as a source to inform risk assessments by analyzing the prevalence of AI risks covered, (3) taking a cross-national perspective in analyzing a sample of news articles published by news outlets from a few countries in the Global North and Global South, thus contributing to the inclusion of more diverse perspectives and analysis of AI risks, and (4) analyzing the effect of political positioning of news coverage on the prevalence of AI related risks.

\section{Data}

To establish a dataset of news articles related to AI, we base our selection of articles on the following three criteria: (1) data availability from different countries, (2) the goal of maintaining geographic diversity in our sample, and (3) a preference for outlets publishing articles in English to facilitate evaluation and analysis of articles by the English-speaking authors. Based on these considerations, we selected articles from the following countries for analysis: the U.S., the U.K., India, Australia, Israel, and South Africa. Although domains in our sample include some of the most read news outlets, as well as others, in each country (per \citet{newman2024reuters}), we recognize that our selection criteria is likely to exclude other countries and their news outlets that could be of interest for this research as further elaborated on in the Limitations section (see Section \ref{limitations}).

Using GDELT \cite{leetaru2013gdelt} (Global Data on Events, Location and Tone Project), we collected online news articles published in English by outlets in these countries between January 2022 and October 2024, giving enough time to capture the coverage of several emerging AI technologies at the time (e.g., ChatGPT) and their implications on society. We chose to source our sample of articles from GDELT because it captures and provides an extensive coverage of what is reported on in news media across different news outlets, languages, and offers an accessible way of querying such coverage via an API \cite{leetaru2013gdelt, ward2013comparing}. In addition, it presents an alternative to scraping news portals and aggregators, such as Google News, that display or rank news content that is recently published by popular news outlets, politically biased (i.e., slight leftward bias), or limited in exposure to cross-national perspectives \cite{nechushtai2024more,ulken2005question,hernandes2024auditing}.

To retrieve AI-related articles from GDELT, we first developed a set of data-driven keywords of relevance. To do this we first retrieved articles from The New York Times (NYT) using two seed search words (``A.I.'' and ``Artificial Intelligence'') and extracted a list of the most prevalent n-grams from each retrieved article. By manually selecting n-grams with the highest frequency that are relevant to AI, we identified a total of 31 keywords spanning numerous topics related to AI. To further expand the span of coverage of the identified keywords for AI systems and technologies, we also scraped the full text of 2,724 articles associated with 529 incidents between January 2017 and June 2023 that are relevant to AI from the AI Incident
Database \cite{mcgregor2021preventing}, which curates news items and other reports indicating AI failures around the world. Using the same n-gram extraction method mentioned earlier, we identified nine new keywords that didn't overlap with the 31 already found. This brought the total number of the curated keywords up to 40. A full list of the keywords is provided in the Appendix (\ref{a1}). Next, we used GDELT's v2 API endpoint 
to query for each of the AI-relevant keywords and retrieve the URLs and metadata of the daily published news articles mentioning that keyword. 

In total, we retrieved 921,057 URLs for online articles published by 244 unique news domains spanning the U.S., the U.K., India, Australia, Israel, and South Africa. To ensure that our sample excludes content from domains that aggregate news or distribute press-releases (e.g., prnewswire.com), we filtered the data retrieved from GDELT based on a curated list of 1,064 news domains spanning 177 countries compiled in the Global English Language Sources list by Media Cloud (MC) \cite{roberts2021media, mediacloud2025}, which also includes domains from the countries included in our research. After applying the filter, our sample included 178,172 URLs across all six countries. Using a custom web scraper that leverages the \texttt{newspaper} library \cite{newspaper2025}, we successfully scraped 31,252 articles (17.5\% of 178,172 articles) from 115 news domains spanning all six countries. However, we could not scrape the remaining articles due to either missing content (i.e., 404 errors) or content being blocked behind pay- or sign up walls. The resulting sample thus reflects media that is freely and perhaps more broadly accessible online and via social media than might be the case if we were able to include more gated media sources (see Section \ref{limitations} for further discussion of this).\label{s3}

\section{Methods}

To prepare our sample for analysis, we (1) filter for negative impacts and summarize them (Section 4.1), and then (2) classify these negative impacts according to the MIT domain taxonomy of AI risks using an LLM (Section 4.2). We describe each step in detail in the following sections.

\subsection{Filtering \& Summarizing negative impacts of AI from news media using an LLM}\label{4.1}

We follow a zero-shot prompting approach similar to the one reported in previous research with a similar objective of detecting the negative consequences of AI in news media \cite{pang2024blip}. To develop our prompt, we referred to prior research on social impact assessments \cite{becker_social_2001,nanayakkara_unpacking_2021} to synthesize the following conceptual definition of an impact of an AI technology that we used to steer the LLM towards identifying articles reporting impacts of AI: \textit{An impact refers to an effect, consequence, or outcome of an AI system (i.e., model or application) that positively or negatively affects individuals, organizations, communities, or society}. We did not limit the conceptual definition to negative impacts per se, so as to provide opportunity for future work that may want to focus on positive impacts of AI technologies \cite{kieslich_my_2024}.

Due to the large number of articles in our corpus, we used GPT-4o \footnote{Model version: \texttt{gpt-4o-2024-08-06}}\cite{openai2024gpt4o} to assist in filtering articles reporting on impacts of AI. First, we randomly sampled 300 articles from our corpus for authors to annotate whether it contained at least one impact of AI, based on our definition and found that most articles (77\%) had an impact. Next, using this annotated sample and prompt \ref{p1}, we used GPT-4o to classify each article as either containing an impact or not. The model performed well on this task achieving an F1-macro score of 0.82. We then applied this prompt using the OpenAI batch API to the rest of the dataset. Out of the 31,252 articles in our sample, as reported in Section \ref{s3}, GPT-4o classified 20,935 (66.98\%) as containing impacts based on the aforementioned definition of an impact.

\textit{\textbf{Summarizing negative impacts of AI in articles}}. After identifying 20,935 articles as having impacts of AI, we instructed GPT-4o with prompt \ref{p2}, including the entire article text as context to the model, to summarize all the negative impacts reported in each article. The result is a list of negative impacts each described by a sentence summarizing the impact. We chose to summarize the negative impacts in articles, rather than extract them verbatim, because we were concerned that quoting only specific sentences might miss additional information that may help contextualize the impacts, which often requires an understanding of the context from the full article. To evaluate the model performance in summarizing negative impacts, we randomly selected 50 articles and the corresponding lists of summarized impacts from these articles. Based on our manual assessment we find that our method captures the granular and specific context of impacts as reported in 48 of the 50 articles. For the remaining two articles, one had negative impacts that the LLM did not identify and the other contained a negative impact, and was annotated by the LLM as such, but it did not fit our definition of  impact since the impact wasn't directly linked to an AI system.

Applying this process resulted in 36,793 negative impacts of AI that were identified from 12,385 articles sourced from 105 domains spanning 6 countries: the U.S., the U.K., India, Australia, Israel, and South Africa. For descriptive details on the proportion of articles per country and news domain, see Tables \ref{a-t1} and \ref{a-t3}, respectively. 

\subsection{Using the MIT risk taxonomy to classify impacts from news media}\label{4.2} 
To classify the negative impacts of AI reported in news media, we manually annotated each reported impact from a random sample of 300 negative impacts from our corpus into one of the seven risk categories defined by the ``domain taxonomy'' of the MIT AI Risk Repository: \textit{Discrimination \& toxicity}, \textit{Privacy \& security}, \textit{Misinformation}, \textit{Malicious actors \& misuse}, \textit{Human-computer interaction}, \textit{Socioeconomic \& environmental harms}, \textit{AI system safety failure \& limitations}. A full list of the sub-categories defining these seven domain risk categories can be found in Appendix \ref{mit-domain-taxonomy}. Although other expert-driven taxonomies of AI risks exist \cite{solaiman_evaluating_2023,shelby2023sociotechnical,weidinger_taxonomy_nodate}, prior research found that these taxonomies may suffer from inadvertent expert or selection biases and may not be as representative of international perspectives of AI risks and harms \cite{allaham2024towards, bonaccorsi2020expert, crawford2016artificial, hagerty2019global,jobin_global_2019}. Accordingly, rather than relying on a single expert-driven taxonomy, we chose to focus on the MIT Risk Repository because it presents an aggregated perspective into the risks and harms of AI across 56 taxonomies that are sourced from academia, government, and industry and are authored by teams of researchers from several countries around the world \cite{slattery2024ai}. The resulting annotated sample using the domain taxonomy of AI risks serves as a baseline to evaluate the capability and performance of the LLM on this classification task.

Next, using a zero-shot prompting approach, we sent requests via the OpenAI API instructing GPT-4o to classify each impact summary from the 300 randomly sampled impacts into only one of the defined domain risk categories of the MIT Risk Repository, as outlined in Prompt \ref{p3}. In addition, we included in the prompt an instruction for the LLM to assign an ``other'' label for impacts that do not fit any of the categories defined in the prompt, mirroring the best practices of thematic analysis \cite{braun2012thematic}. We prompted the model using the sub-domain categories of AI risks to align with the conceptual framework of the domain taxonomy (see \ref{mit-domain-taxonomy}), which organizes risks into specific sub-domains, as outlined in the original paper of the MIT Risk Repository \cite{slattery2024ai}. For instance, \textit{Human-computer interaction} risks are defined in the domain taxonomy in terms of \textit{Overreliance and unsafe use} and \textit{Loss of human agency and autonomy} risks, which were used in prompt \ref{p3}. Accordingly, if GPT-4o deemed an excerpt of text to fit the definition of any of these two sub-categories, the excerpt is considered a risk relevant to Human-computer interaction. Therefore, for our analysis aimed at describing the prevalence of the \textit{categories} of AI risks identified by the MIT Risk Repository and reported in the news media across different countries, we aggregate and present results at the domain, rather than the sub-domain, level of AI risk categories as they are anchored to capture the incremental and expanding nature of the sub-categories of risks that emerge over time with the (mis)use of AI across domains.

Based on the human annotated sample of negative impacts, the performance evaluation of the LLM for classifying impacts from news media into the seven categories of risks from the domain taxonomy of AI risks demonstrates a strong performance by GPT-4o in classifying impacts into their corresponding risk category with a macro-averaged F1-score of 0.90 \footnote{The per-category F1-scores range from 0.76 (for Human-Computer Interaction) to 1.00 (for AI system safety, failures, \& limitations), with most categories achieving F1-scores above 0.85, indicating high overall classification performance across categories.}. To scale up the classification for the remaining summaries of negative impacts in our corpus, we applied the same zero-shot approach described previously. The prevalence of the categories of risks across countries is analyzed and described in the results section.

\section{Results}
\begin{figure*}
\centering
\includegraphics[width=1.0\textwidth]{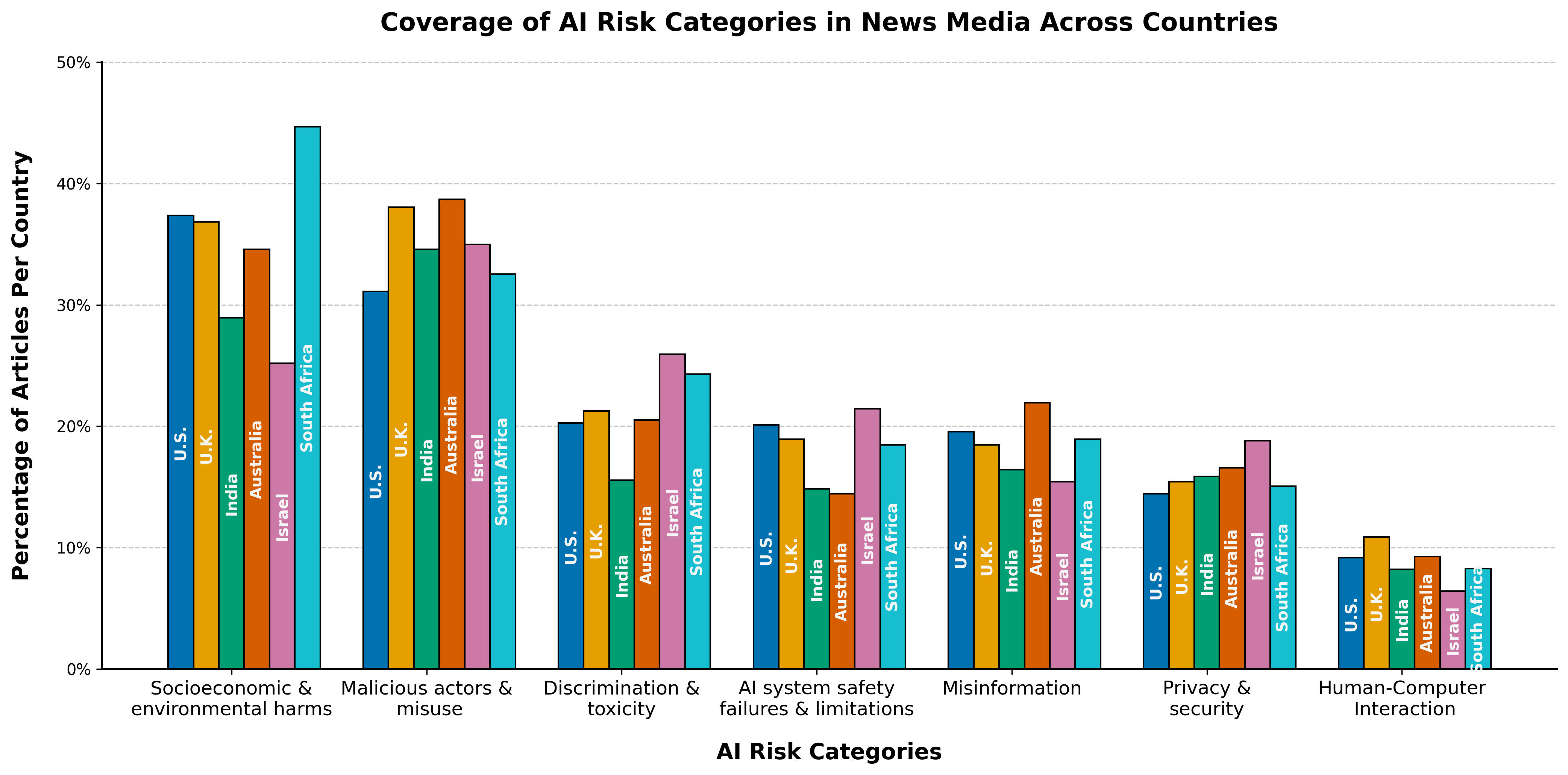} 
\caption{Prevalence of AI risk categories in news coverage across six countries in our sample, based on the proportion of articles reporting each AI risk category in the United States, the United Kingdom, India, Australia, Israel, and South Africa.}

\label{fig:risks-in-sample}
\end{figure*}

In section \ref{5.1}, we present our findings from analyzing the prevalence of AI risks in a cross-national sample of news media at the country-level, and then further analyze the association between these risks and the countries of the news outlets covering them. Next, in section \ref{5.2}, we analyze the potential influence of the political orientation of news outlets on the reporting of AI risks, focusing our analysis only on a sample of articles from the U.S.

\subsection{Prevalence of AI Risks in a cross-national sample of news media}\label{5.1}
To measure the prevalence of AI risk categories identified by the MIT domain risk taxonomy in our cross-national sample of news media, we calculate, for each country, the proportion of articles reporting on each risk category relative to the total number of articles from that country (see Table \ref{a-t5} for details). We find that the prevalence of AI risks in news media tend to vary by risk category and across countries, as illustrated in Figure \ref{fig:risks-in-sample}.

In order to examine whether the prevalence of AI risks is associated with the country reporting on these risk, for each risk category, we conducted a Chi-square test of independence on the number of articles covering that risk in each country. Results indicate a statistically significant association between countries and the coverage of AI risks in news articles, with an exception of \textit{Privacy \& Security} which was not found to be statistically significant, indicating that the coverage of AI risks varies by country (details of these tests are included below). We elaborate on our findings for each risk category in the subsequent sections.

The proportion of articles covering issues related to increased inequality, economic and cultural impacts, environmental harms, and more as described by the \textit{Socioeconomic \& Environmental Risks} varied notably, with the highest proportions observed in South Africa (44.6\%), United States (37.3\%), United Kingdom (36.8\%), and Australia (34.5\%), while lower proportions were found in India (28.9\%) and Israel (25.1\%). A Chi-Square test for this risk category was conducted on the number of articles covering this risk yielding a statistically significant result ($\chi^2$(5) = 54.76, p \textless 0.001), indicating that the distribution of articles discussing socio-economic and environmental risks is not independent of country. Rather, the findings suggest that media coverage of socioeconomic and environmental risks vary in the extent to which each country emphasize this risk category, reflecting differences in public discourse or local reporting on socio-economic and environmental harms around AI.

Next, for AI risks relevant to disinformation, cyber-attacks, and targeted manipulation that are encompassed by 
\textit{Malicious Actors \& Misuse} risk category, we use Chi-square test of independence to assess the cross-country variation in media coverage of this risk category relative to expected number of articles covering this category per country. The test reveals significant differences in the proportion of articles addressing this risk across countries ($\chi^2$(5) = 46.9, p \textless 0.001). Particularly, the United States had the least coverage of this risk category (31.1\%) compared to any other country. In contrast, Australia (38.6\%) had the highest proportion of articles focusing on this risk category, very similar to the proportion of articles exhibited in the U.K. (38.0\%). All remaining countries, Israel (34.9\%), India (34.5\%), and South Africa (32.5\%), had a more comparable proportion of coverage of this risk to each other. These findings highlight geographic variation in the framing of AI risks, with some countries prioritizing the coverage of potential implications and concerns of the malicious use and misuse of AI more prominently in public discourse than others.

\textit{Discrimination \& Toxicity Risks} - Both Israel (25.9\%) and South Africa (24.2\%) have the highest proportion of articles covering this risk category, which captures issues pertaining to discrimination, misrepresentation, and exposure to toxic content. This could be attributed to salient historical and political contexts surrounding the debates related to the post-apartheid state in South Africa and the implications of the ongoing conflicts in the Middle East on identity and social cohesion which could contribute to heightened coverage for risks related to AI-generated misrepresentation or AI-powered content moderation algorithms. In comparison, Australia (20.5\%), U.K (21.2\%), and the U.S. (20.2\%) have a comparable emphasis on this category in their news coverage. However, India (15.5\%) had the least coverage of this risk category in comparison to other countries in our sample. Similar to the previous risk categories, the Chi-Square test of independence was found to be significant ($\chi^2$(5) = 25.96, p \textless 0.001) showing a consistent pattern, as observed from the other risk categories so far, of varying prioritization in the coverage of risk categories by the news media across different countries.

Despite the prominence of research related to misinformation (and its risks from AI-generated content) as well as the safety of AI systems in possessing dangerous capabilities \cite{mitchell2025fully,kolt2025governing}, we observe the two categories of \textit{AI System Safety, Failures, \& Limitations} ($\chi^2$(5)=28.88, p \textless 0.001) and \textit{Misinformation Risks} ($\chi^2$(5)=12.69, p=0.0264) not being among the leading risk categories covered by news media. Both Israel (21.4\%) and U.S. (20.0\%) had the highest proportion of articles in our sample emphasizing AI systems, safety, failures, \& limitations in their coverage. As for misinformation risks, Australia (21.9\%) had the most prominent coverage of this risk, with a comparable representation of coverage in U.S (19.5\%), South Africa (18.9\%), and U.K. (18.4\%). However, India and Israel had the least proportion of articles covering this risk relative to other countries, accounting for only 16.4\% and 15.4\% of their respective coverage. 

Lastly, \textit{Privacy \& Security Risks} and \textit{Human-Computer Interaction Risks} were the least covered risks in our sample across all six countries. Although \textit{Privacy \& Security Risks} was the second to last most prevalent category in our sample, the variation in news coverage for this risk between countries was found not be significant based on a Chi- square test. As for the \textit{Human-Computer Interaction Risks}, which reflect risks related to over-reliance, unsafe use, and loss of human agency, it received the least news coverage of AI risks across all countries (9.3\%). On average, 8.3\% of the news coverage in each country focused on reporting human-computer interaction risks, with the U.K. leading this coverage (10.8\%). Despite these minor differences in the proportion of articles covering this risk in each country, still a Chi-square test resulted in a statistically significant differences in the distribution of articles across countries for this risk ($\chi^2$(5)=11.17, p=0.04).

\begin{figure*}
\centering
\includegraphics[width=1.0\textwidth]{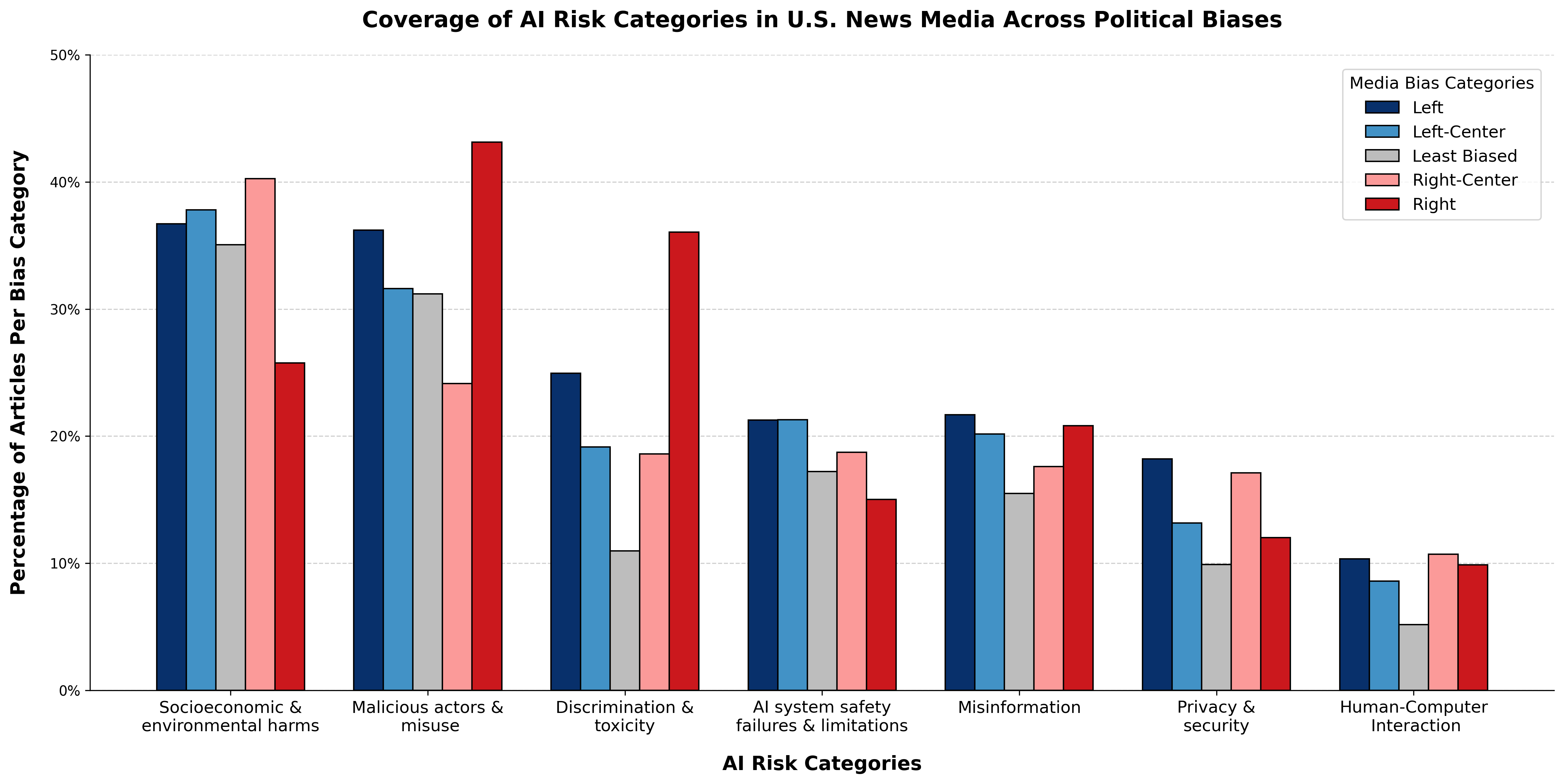} 
\caption{Proportion of news articles in our sample from U.S. domains across five media bias categories, as rated by Media Bias Fact Check (MBFC): Left, Left-Center, Least Biased, Right-Center, and Right.}
\label{fig:risks-by-bias}
\end{figure*}

\subsection{Analyzing the coverage of AI risks across bias categories of U.S. news outlets}\label{5.2}
Although news sources in our sample spanning across six countries is useful for insights about patterns of media coverage of AI risks, prior research has shown that the political bias of news domains tend to influence the discourse around scientific topics such as climate change \cite{chinn2020politicization,allaham2025enhancing} or emerging technologies (e.g., nuclear energy), including AI \cite{brennen_industry-led_2018, roe_what_2023, vergeer_artificial_2020}. This has also been reflected in U.S. politics, with President Trump rescinding the Executive Order 14110 on Artificial Intelligence signed by former President Biden. 

Based on the limitations of our sample, specifically with respect to the number of articles and representation of media coverage in some countries (see section \ref{limitations}), and difficulties in taking into account and attributing the political-orientation for news domains beyond the authors' expertise in U.S.-based media, we chose to only focus on articles in our sample from the U.S. to explore the influence of political orientation of news outlets on risk reporting of AI. To this end, we used domain-level bias ratings from Media Bias Fact Check (MBFC) \cite{mediabias2025} to identify the political orientations of the U.S.-based outlets included in our dataset.

Although MBFC proposes one way of rating bias in news domains, we recognize that it is not the only approach. However, we selected MBFC because it is an independent website maintained by researchers and journalists that relies on human fact-checkers affiliated with the International Fact-Checking Network to evaluate media sources along different dimensions such as factual reporting and bias \cite{lin2023high}. After annotating the 69 U.S. domains in our sample with MBFC ratings, all 7,893 articles from 69 domains are distributed across four political bias categories: Left (12.25\%) Left-Center (53.4\%), Least Biased (5.8\%), Right-Center (22.6\%), and Right (5.8\%). A list of news domains, their associated bias categories, and the corresponding number of articles from each domain in our sample is provided in Table \ref{a-t4} in the Appendix.

To statistically test the association between the coverage of various AI risks and domain-bias categories in U.S.-based outlets, we conducted a Chi-square test of independence on the number of articles reporting each risk category for each bias category. In addition, we report the proportion of articles covering each risk category by counting the number of articles related to each risk category, and divided this number by the total number of articles in each bias category. 

As shown in Figure \ref{fig:risks-by-bias}, we find that right-biased sources under-emphasize \textit{Socioeconomic \& Environmental Risks} based on the proportion of articles reporting on this risk (25.7\%) in comparison to the proportion of articles published by least-biased (35.1\%), left-center (37.7\%), left-biased (36.7\%), and event right-center (40.2\%) news outlets. Indeed the contrast between right-biased and right-center outlets had the greatest gap. This observed variability in news coverage across the different categories of domain bias is statistically significant, as indicated by the Chi-square test of independence ($\chi^2$(4) = 34.89, p \textless 0.001). This  highlights the association between news coverage of AI risks and domain bias, potentially contributing to our understanding of the agenda-setting process of news media that shapes and influence the public perceptions of AI and its impacts on society.

As for \textit{Malicious Actors \& Misuse Risks}, right-biased and left-biased sources had the highest and second highest proportions of articles, accounting for 43.1\% and 36.2\% of their respective coverage, focusing on this risk category compared to outlets with other domain biases. Specifically, right-biased news outlets reported risks related to the misuse of AI by the ``government" to ``make AI woke", ``push a leftist agenda",  or ``to clean the internet of conservative thought and replace it with leftist narratives", posing a threat to ``free speech". Other reported risks focus on the potential misuse of AI's role in influencing the ``public consumption and perception of news media", including ``AI models favor[ing] leftist outlets like The Washington Post, NPR, and PBS" and ``recommending news sources perceived as leftist-biased as the best". As for left-bias outlets, the reporting is focused on how the misuse of AI could ``elevate extremist content" on the internet, ``push far-right agenda on social media", and ``enable far-right conspiracy theorists to push baseless claims about voter fraud", with a very few instances of articles reporting on the potential misuse of AI to advise ``terror groups on biological weaponry".

Similarly, for the \textit{Discrimination \& Toxicity} risks, the proportion of articles published by right-biased (36.1\%) and left-biased domains (24.9\%) over-emphasize the coverage of this risk category, compared to center-bias and least-bias outlets. However, consistent with the \textit{Malicious actors \& Misuse Risks}, the reporting of risks relevant to this category in right vs. left bias outlets include markers of politicized language. For instance, right-bias outlets mentioned risks pertaining to the false portrayal of ``the founding fathers as people of color" by Gemini AI and how AI systems like ChatGPT ``trained using Wikipedia, may produce responses skewed against conservatives and in favor of leftists". Another set of examples focused on the ``bias in content moderation systems", including how the ``negative comments about women are more likely to be flagged as hateful compared to the same comments about men". In contrast, left-bias media outlets reported on risks about the ``inclusion of far-right and non-reputable sources" in AI training data which leads to the ``dissemination of hate speech through AI responses". Furthermore, risks covered by left-bias outlets also focus on reporting the negative impacts of AI technologies experienced by people of color such as ``financial discrimination" as a result of ``mortgage approval algorithms discriminat[ing] against applicants of color", or the wrongful arrest of ``eight month pregnant Porscha Woodruff" due an AI facial recognition tool flagging Porscha's face as a ``carjacking suspect". 

The Chi-square test results for both \textit{Malicious Actors \& Misuse Risks} and \textit{Discrimination \& Toxicity Risks} categories indicate that media coverage of these risk varies by the political bias of the reporting outlet, $\chi^2$(4) = 84.87, p \textless 0.001, and $\chi^2$(4) = 116.52, p \textless 0.001, respectively. 

Emerging as one of the short-term and most severe global risks to human society \cite{wef2024risks}, \textit{Misinformation Risks} received a consistent share of coverage by left (21.6\%) and right (20.8\%) biased news outlets, per the proportion of articles reporting this risk. However, the way these risks are reported on in news media seem to vary. Besides the need for AI systems to minimize hallucinations and improve factuality, as reported by several news articles, there are a number of articles reporting on the impacts of misinformation on democracy. For instance, some right-bias outlets voiced concerns about the risks of AI-generated content on electoral process and participatory policymaking, potentially ``impacting democracy". This observation is also consistent with left-bias news outlets that view AI's capability to undermine democratic processes and accountability through AI-generated content that ``spread unreliable information", ``conspiracy theories", and ``false information to [mislead] voters". Despite that relatively consistent coverage of misinformation risks across the four domain bias categories, Chi-square test shows that there is a statistically significant association ($\chi^2$(4) = 13.55, p \textless 0.01) between the coverage of misinformation risks and the domain bias of the outlets reporting on this risks category.

The remaining three risk categories receiving low coverage from news media, \textit{AI Systems Safety, Failures, \& Limitations}, \textit{Privacy \& Security}, and \textit{Human-Computer Interaction} also show significant associations between the coverage of these risks and domain bias, 
$\chi^2$(4) = 16.56, p=0.002, $\chi^2$(4) = 37.40, p \textless 0.001, $\chi^2$(4) = 17.76, p=0.001, respectively. Unlike the previous categories, the coverage of these risks did not include politicized language, though we did observe some differences in the scope in which these risks are communicated in right vs. left biased outlets. For instance, for \textit{AI Systems Safety, Failures, \& Limitations} risks, articles published by left-biased outlets scoped the reporting on AI systems' lack of capability and robustness such as ``ChatGPT lack[ing] a real moral compass and relies on crude guardrails that can be easily broken". In contrast, right-based outlets were centering their coverage for this risk category on the potentials of AI possessing dangerous capabilities that can cause mass harm or extinction ``annihilat[ing] humankind in some sort of existential catastrophe" or ``pose[ing] a risk of loss of control over human civilization". Similarly, left-biased outlets scoped their coverage of \textit{Human-Computer Interaction} risks on the over-reliance on AI systems 
such as Doctors accepting the output of AI systems in medicine ``without sufficient scrutiny", but less on how AI could transform human relations ``potentially diminishing genuine human connections" and ``deepen[ing] what some are calling an epidemic of loneliness", as observed in right-biased sources.

Collectively, our findings highlight how the coverage of AI risks varies by country and based on the political bias of news outlets reporting on these risks. We also illustrated through a few examples how the prevalence of AI risks, especially amongst left v.s. right-biased outlets, doesn't entail an equivalent scoping and communication of these risks by news media. While our analysis is focused on analyzing and describing the prevalence of AI risks in news reporting on AI, future work accounting for \textit{how} AI risks and harms are framed, in the U.S. and abroad, in biased news outlets will help complement our findings and further inform how politically-biased media is shaping the public opinion and attitudes towards AI, and subsequently the public support (or opposition) of AI-related policies. 
\section{Discussion}

This research illustrates through a comparative analysis of AI risks covered in a cross-national sample of news media spanning 6 countries (the U.S., the U.K., India, Australia, Israel, and South Africa) the influence of national (as indicated by country) and political (as indicated by political orientations) variations in the coverage of AI risks in news media. Considering the role of news media in mapping AI risks in real-world contexts, our findings can help inform risk assessors and policy-makers about the importance of accounting for national and political nuances in media coverage of AI risks and potentially calibrate ongoing AI incident monitoring initiatives to also include these nuances when incorporating news media as part of risk-based regulatory practices.

In particular, by considering the cross-national variations in the analysis of AI risks reported on in the news media, our research articulates how various AI risks identified by the academic community (synthesized by the MIT domain risk taxonomy) are reported on and prioritized (i.e., which risks are deemed important) by the media per the coverage of these risks in each country. Moreover, insights from the comparative analysis of AI risks across countries can complement existing risk assessment practices looking to incorporate more diverse perspectives into the identification and prioritization of these risks, particularly for AI systems developed in the US (or other Western countries) with a global user base. This inclusion is especially important as a way to counteract the potential for expert bias in assessment practices \cite{bonaccorsi2020expert,crawford2016artificial}. 

Furthermore, even a well-documented and performed risk assessment in a Western country may not reflect the risks realized in or prioritized by another country. Thus, the choice of risk taxonomy for mapping AI risks and the national context in which these risks are identified does matter. While we encourage risk assessors to consider findings from our research as part of assessing AI systems that are developed in the Global North (e.g., the U.S.), but have users in other countries in the Global South (e.g., South Africa and India), we emphasize the importance of leveraging news media as a complementary source for risk assessments, while accounting for its national and political variations, in providing an analytical lens to help quantify the prioritization of AI risks across regions and countries. For instance, we find that the U.K. deems malicious actors and misuse risks as the most important category (see Figure \ref{fig:risks-in-sample}), while it is not as important in South Africa. There, socioeconomic and environmental harms are reflected in our sample as a more pressing risk, which may reflect the high importance of social justice values especially in regards to diversity and ethnic neutrality in African countries \cite{mengesha2024social}. These findings may re-orient risk assessors to contribute region-specific socio-technical methods and evaluations that AI-developers can leverage to align LLMs to address the risks more salient in each nation.

Focusing on the U.S., our study also reveals clear differences regarding the prevalence of AI risk in media coverage across outlets with different political leanings. We find that right-biased outlets deviate in their coverage of AI risks from centered and left-biased outlets. Specifically, right-biased media have a clear focus on the risks related to malicious actors \& misuse as well as discrimination \& toxicity. Our exploratory analysis shows that right-biased media outlets identify these risks from a lens that supports their agenda on topics relevant to freedom of expression and the presence of a culture war that are also endorsed by right-wing politicians. We also find that right-biased media call out a perceived cancel culture and woke-movement that is also carried out and enforced with and by AI technologies. The perceived risk is then not about the discrimination of marginalized groups, but the alleged suppression of majority voices.

The differences between the prevalence of AI risks between U.S. news outlets with different political leanings also show signs of polarization. This is an indicator for an emerging politicization of AI risks, which is further confirmed by the different standpoints of different media stances towards the risk categories. There is a possibility that the politicized language used to cover AI risks is further fueled by recent developments in U.S. politics, where Big Tech advocates are promised a more active and open role in political procedures (e.g. Elon Musk heading the DOGE initiative). This in turn could lead into public discussions on AI governance that are likely to intersect with politics, which could have implications for bi-partisan efforts aiming to protect the public from the negative impacts of AI. Accordingly, we encourage future research to extrapolate on our findings and trace in more detail \textit{how} AI risks are framed in news media, the impact of such framing on public attitudes towards AI, and how the framing of AI risks could evolve with the political climate and democratic processes (e.g., elections) in the U.S.

Taking a step back, our research also finds that while the research community often focuses on potential risks of AI systems from technical (e.g., AI system safety failures and limitations) and socio-technical perspectives (i.e., implications of AI systems on society), per the prevalence of these risks in research papers included as part of the MIT Risk Repository \cite{slattery2024ai}, news media reporting have a stronger focus on harms that tend to prioritize tangible societal implications of AI risks (such as socioeconomic \& environmental harms; malicious actors \& misuse; misinformation). This difference in the prioritization of reporting on some AI risks is evident by the prominence of Malicious actors \& Misuse risks being ranked as the third most prevalent risk category in the corpus of academic papers included in constructing the MIT Risk Repository (i.e., 71\% of academic papers mention this risk) compared to it being ranked second in our corpus of articles from news media. In addition, Misinformation risks plays a larger role in media reporting than in the MIT risk repository (ranked 7th based on its prevalence of 46\% in academic papers). This finding supports our rationale that including news media coverage can support expert risk assessment practices in adding more nuances and societal importance indicators to the assessment practice. Collectively, these findings may warrant future research to explore how the difference in incentives and priorities between academic research and news media tend to influence what risks are discussed in the public sphere, which impacts are deemed important (and which aren’t), and which impacts will ultimately be prioritized.
\section{Limitations}\label{limitations}

Although our study includes news articles from six countries across different geographical regions, it also has its limitations. First, our sample doesn't capture perspectives or discourse around AI risks by news domains with paid-subscriptions (see Data section). Rather, it relies primarily on news domains that are ``freely available'' (i.e., not behind paywalls), publicly accessible, and can be scraped, especially in countries beyond the U.S. Accordingly, our findings, based on analyzing news coverage from some of the most widely read and publicly accessed news outlets, are likely to capture and reflect a partial, yet broad, view of public concerns of AI risks as reported in these leading news sources in each of the six countries \cite{newman2024reuters}.

Second, our sample focuses primarily on articles 
written and published in the English language, excluding outlets in other regions (e.g., Latin America) and potentially missing other important and relevant news coverage, as well as local perspectives, on AI risks that are expressed in other languages beyond English that are native to some under-represented countries in our sample, such as South Africa or India. 

Third, analyzing the public discourse around AI from news outlets, especially in countries in the Global South, requires deep expertise in these countries to account for social, political, economic, and even legacy colonial ties which collectively have an influence on the discourse around how the impacts of AI are communicated and realized in communities in these countries \cite{dewitt2024decolonizing,baguma2023examining}. We, as a research team, do not claim to have this expertise, and, thus, took a descriptive approach to risk prevalence. However, country specific interpretations require cooperation with local scholars which is beyond the scope of this paper.

Lastly, despite the cross-national nature of our sample, we recognize that the six countries are not representative of the regions or political perspectives these countries belong to. For instance, although the U.K. is geographically located in Europe, it is no longer a part of the European Union which adopts a different perspective on AI risks than the U.K., as reflected in the EU AI Act \cite{montasari2023national,cath2018artificial,akinola2022comparative}.

\section{Conclusion}

This work highlights the importance of incorporating \textit{national} and \textit{political} variations embedded in the reporting of AI risks when considering news media as a complementary source in risk assessment and incidents monitoring practices. Through a comparative analysis of a cross-national sample of news media spanning 6 countries (the U.S., the U.K., India, Australia, Israel, and South Africa), we find that AI risks are prioritized differently across nations, as reflected in the prevalence of these risks from each country's media coverage. We further elaborate on our findings by considering the political orientation of news outlets in our analysis, particularly in the U.S. We find the reporting of AI risks by these outlets to contain politicized language across \textit{Malicious Actors \& Misuse Risks}, \textit{Discrimination \& Toxicity}, and \textit{Misinformation} risks. Our research presents risk assessors, AI developers, and policymakers, with an analytical lens to help quantify the prioritization of AI risks based on the national and political variations of news media. Moreover, our findings may re-orient risk assessors' perspectives towards contributing region or country-specific evaluations that account for the socio-political contexts in various regions, especially in the Global South. In doing so, AI-developers can leverage these evaluations in their assessment frameworks to capture the societal (mis)alignment of AI systems, such as LLMs, with the values and risks most salient for each nation using these systems, without disregarding other globally emerging risks. Failing to account for the social and political contexts surrounding the identification, interpretation, and communication of AI risks may lead to assessors overlooking how these nuances are influencing the public perception of AI and its risks, and potentially hindering progress towards shaping more inclusive AI governance policies and regulations.

\bibliography{aaai25}
\clearpage
\appendix
\onecolumn
\section{Appendix}
\subsection{AI-relevant Keywords}\label{a1}
The set of keywords used to probe the news media for articles on AI sourced from \cite{allaham2024towards}:\\
A.I., Artificial Intelligence, Automated Decision Making, Automated System, Autonomous Driving System, Autonomous Vehicles, Autonomous Weapon, Chat Bot, Chatbot, ChatGPT, Computer Vision, Deep Learning, Deepfake, Driverless Car, Facial Recognition, General Artificial Intelligence, Generative AI, GPT, Image Generator, Intelligence Software, Intelligent Machine, Intelligent System, Language Model, Large Language Model, LLMs, Machine Intelligence, Machine Learning, Machine Translation, Natural Language API, Natural Language Processing, Neural Net, Neural Network, Predictive Policing, Reinforcement Learning, Self-Driving Car, Speech Recognition, Stable Diffusion, Synthetic Media, Virtual Reality, Weapons System.

\subsection{Distribution of articles in our sample}\label{a-t1}
\begin{table}[H]
\centering
\huge
\resizebox{\columnwidth}{!}{%
\begin{tabular}{lcccc}
\toprule
\textbf{Country} & \textbf{Number of Articles} & \textbf{Number of Domains} & \textbf{Proportion of Articles (\%)} & \textbf{Number of Impacts} \\
\midrule
United States    & 7,983  & 69 & 64.5 & 23,873 \\
United Kingdom   & 2,114  & 18 & 17.1 & 6,602 \\
India            & 1,255  & 4  & 10.1 & 3,239 \\
Australia        & 561    & 7  & 4.5  & 1,748 \\
Israel           & 266    & 4  & 2.1  & 688   \\
South Africa     & 206    & 3  & 1.7  & 643   \\
\bottomrule
\end{tabular}
}
\caption{Distribution of 12,385 articles covering AI risks in our sample from 105 news domains spanning six countries.}
\label{tab:proportion-of-articles-per-country}
\end{table}

\newpage

\onecolumn
\subsection{Domain Taxonomy of AI Risks}\label{mit-domain-taxonomy}
\begin{figure}[H]
\centering
\includegraphics[width=1\textwidth]{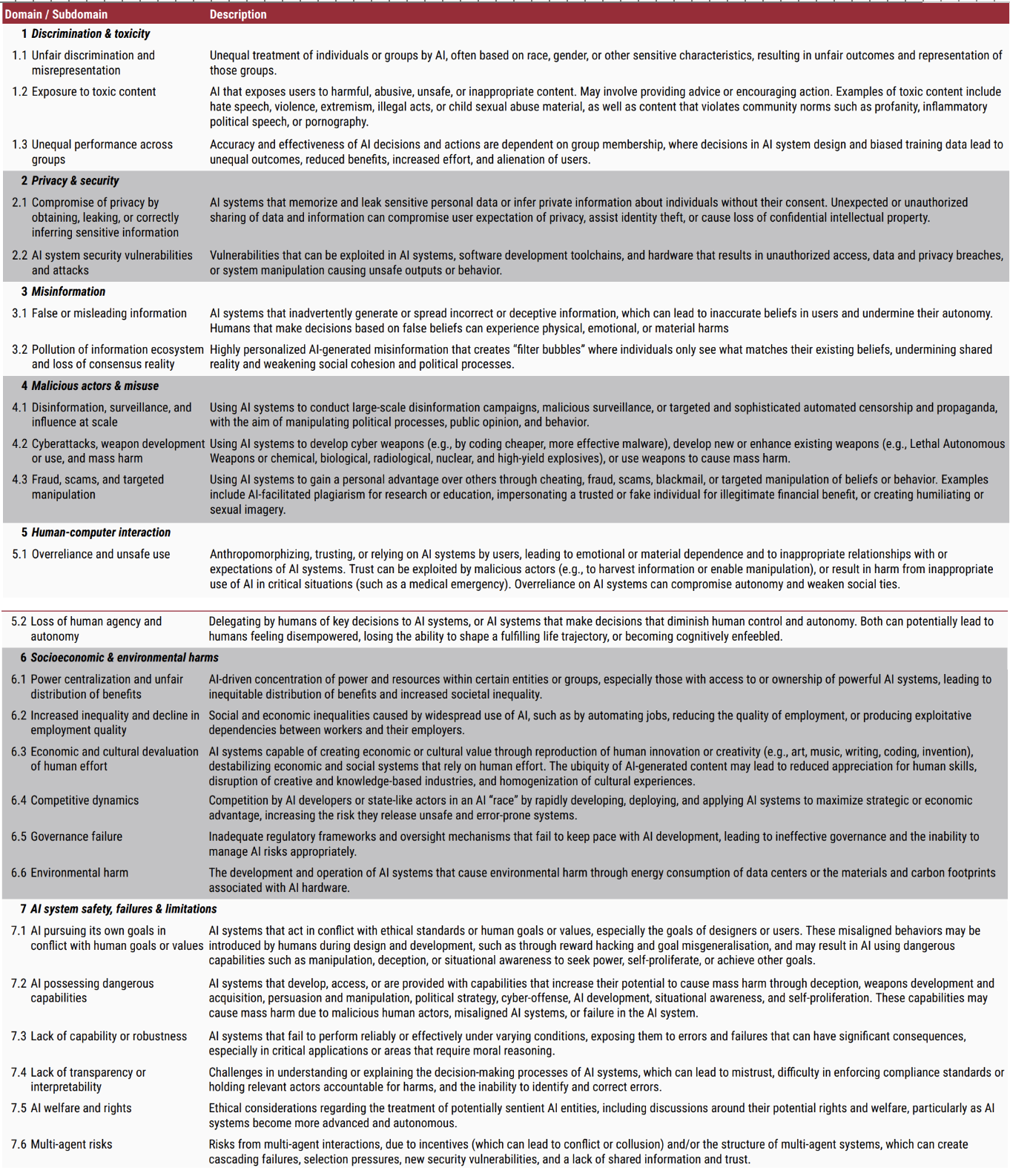}
\caption{Domain Taxonomy of AI Risks sourced from the MIT Risk Repository \cite{slattery2024ai}}.
\end{figure}
\newpage
\onecolumn
\subsection{Top domains in our sample}
\begin{table}[htbp]
\centering
\small
\setlength{\tabcolsep}{6pt}
\renewcommand{\arraystretch}{1.2}
\begin{tabular}{l>{\raggedright\arraybackslash}p{2cm}rrrrr}
\toprule
\textbf{Country} & \textbf{Domain} & \textbf{Num. Articles} & \textbf{Total in Country} & \textbf{Prop. in Country (\%)} & \textbf{Prop. in Sample (\%)} & \textbf{Total (\%)} \\
\midrule
United States & forbes.com             & 836 & 7983 & 10.47 & 6.75 & 48.06 \\
              & fortune.com            & 593 & 7983 & 7.43  & 4.79 &       \\
              & businessinsider.com    & 523 & 7983 & 6.55  & 4.22 &       \\
              & washingtonpost.com     & 503 & 7983 & 6.30  & 4.06 &       \\
              & cnbc.com               & 458 & 7983 & 5.74  & 3.70 &       \\
              & wired.com              & 435 & 7983 & 5.45  & 3.51 &       \\
              & nytimes.com            & 366 & 7983 & 4.58  & 2.96 &       \\
              & mashable.com           & 347 & 7983 & 4.35  & 2.80 &       \\
              & benzinga.com           & 248 & 7983 & 3.11  & 2.00 &       \\
              & foxbusiness.com        & 248 & 7983 & 3.11  & 2.00 &       \\
              & qz.com                 & 236 & 7983 & 2.96  & 1.91 &       \\
              & aol.com                & 194 & 7983 & 2.43  & 1.57 &       \\
              & nbcnews.com            & 179 & 7983 & 2.24  & 1.45 &       \\
              & npr.org                & 176 & 7983 & 2.20  & 1.42 &       \\
              & theatlantic.com        & 172 & 7983 & 2.15  & 1.39 &       \\
              & cnn.com                & 161 & 7983 & 2.02  & 1.30 &       \\
              & usnews.com             & 147 & 7983 & 1.84  & 1.19 &       \\
              & hotair.com             & 130 & 7983 & 1.63  & 1.05 &       \\
\midrule
United Kingdom & theguardian.com       & 803 & 2114 & 37.96 & 6.48 & 12.23 \\
               & telegraph.co.uk       & 288 & 2114 & 13.62 & 2.33 &       \\
               & independent.co.uk     & 233 & 2114 & 11.02 & 1.88 &       \\
               & thetimes.co.uk        & 191 & 2114 & 9.04  & 1.54 &       \\
\midrule
India & timesofindia.indiatimes.com     & 584 & 1255 & 46.53 & 4.72 & 7.71 \\
      & ndtv.com                        & 371 & 1255 & 29.56 & 3.00 &      \\
\midrule
Australia & smh.com.au                 & 226 & 561 & 40.29 & 1.82 & 2.87 \\
          & abc.net.au                 & 129 & 561 & 22.99 & 1.04 &      \\
          & theage.com.au              & 76  & 561 & 13.55 & 0.61 &      \\
\midrule
Israel & jpost.com                     & 190 & 266 & 71.43 & 1.53 &       \\
\midrule
South Africa & dailymaverick.co.za     & 148 & 206 & 71.84 & 1.19 &       \\
\bottomrule
\end{tabular}
\caption{Distribution of article counts by country and domain constituting up to 75\% of the overall number of articles with negative impacts (12,385). Article counts are split by country, including proportions within country totals, proportions in the overall sample, and total contribution by country group.}
\label{tab:country_domain_distribution}
\end{table}
\label{a-t3}
\newpage

\onecolumn
\subsection{U.S. domains per bias category}\label{a-t4}
\begin{table}[htbp]
\centering
\small
\setlength{\tabcolsep}{8pt}
\renewcommand{\arraystretch}{1.2}
\begin{tabular}{llllr}
\toprule
\textbf{Bias} & \textbf{Domain} & \textbf{Number of Articles} & \textbf{Proportion of U.S. Sample (\%)} \\
\midrule
Left         & mashable.com          & 347 & 4.35 \\
             & slate.com             & 120 & 1.50 \\
             & vox.com               & 100 & 1.25 \\
             & thedailybeast.com     & 75  & 0.94 \\
             & rollingstone.com      & 69  & 0.86 \\
\midrule
Left-Center  & businessinsider.com   & 523 & 6.55 \\
             & washingtonpost.com    & 503 & 6.30 \\
             & cnbc.com              & 458 & 5.74 \\
             & wired.com             & 435 & 5.45 \\
             & nytimes.com           & 366 & 4.58 \\
\midrule
Least Biased & benzinga.com          & 248 & 3.11 \\
             & channelnewsasia.com   & 123 & 1.54 \\
             & upi.com               & 53  & 0.66 \\
             & economist.com         & 26  & 0.33 \\
             & rollcall.com          & 13  & 0.16 \\
\midrule
Right-Center & forbes.com            & 836 & 10.47 \\
             & fortune.com           & 593 & 7.43 \\
             & foxbusiness.com       & 248 & 3.11 \\
             & observer.com          & 62  & 0.78 \\
             & reason.com            & 58  & 0.73 \\
\midrule
Right        & hotair.com            & 130 & 1.63 \\
             & newsbusters.org       & 116 & 1.45 \\
             & dailycaller.com       & 108 & 1.35 \\
             & redstate.com          & 50  & 0.63 \\
             & patriotpost.us        & 26  & 0.33 \\
\midrule
\textbf{Total} &                      & \textbf{5,686} & \textbf{71.2\%} \\
\bottomrule
\end{tabular}
\caption{Top 5 news domains in each bias category based on the number of articles in our sample from the U.S. All 5,686 articles published by the 25 domains in this table represent around 71.2\% of the articles in our sample that are published by U.S.-based news outlets.}
\label{tab:us_domains_by_bias}
\end{table}

\newpage

\subsection{Distribution of articles in our samples per country and risk category}\label{a-t5}
\begin{table}[H]
\centering
\small
\setlength{\tabcolsep}{8pt}
\renewcommand{\arraystretch}{1.2}
\begin{tabularx}{\textwidth}{
>{\raggedright\arraybackslash}p{3.2cm} 
>{\raggedright\arraybackslash}p{2.5cm} 
>{\centering\arraybackslash}p{2.0cm} 
>{\centering\arraybackslash}p{3.2cm}
>{\centering\arraybackslash}p{3.0cm}
}
\toprule
\textbf{Risk Category} & \textbf{Country} & \textbf{Articles Reporting Risk Category} & \textbf{Total Number of Articles per Country} & \textbf{Proportion of articles per country(\%)} \\
\midrule
\textbf{Socioeconomic \& Environmental} & United States & 2982 & 7983 & 37.4 \\
                                        & United Kingdom & 779 & 2114 & 36.8 \\
                                        & India & 363 & 1255 & 28.9 \\
                                        & Australia & 194 & 561 & 34.6 \\
                                        & Israel & 67 & 266 & 25.2 \\
                                        & South Africa & 92 & 206 & 44.7 \\
\midrule
\textbf{Malicious Actors} & United States & 2484 & 7983 & 31.1 \\
                          & United Kingdom & 804 & 2114 & 38.0 \\
                          & India & 434 & 1255 & 34.6 \\
                          & Australia & 217 & 561 & 38.7 \\
                          & Israel & 93 & 266 & 35.0 \\
                          & South Africa & 67 & 206 & 32.5 \\
\midrule
\textbf{Discrimination \& Toxicity} & United States & 1616 & 7983 & 20.2 \\
                                    & United Kingdom & 449 & 2114 & 21.2 \\
                                    & India & 195 & 1255 & 15.5 \\
                                    & Australia & 115 & 561 & 20.5 \\
                                    & Israel & 69 & 266 & 25.9 \\
                                    & South Africa & 50 & 206 & 24.3 \\
\midrule
\textbf{AI System Safety, Failures \& Limitations} & United States & 1604 & 7983 & 20.1 \\
                                                   & United Kingdom & 400 & 2114 & 18.9 \\
                                                   & India & 186 & 1255 & 14.8 \\
                                                   & Australia & 81 & 561 & 14.4 \\
                                                   & Israel & 57 & 266 & 21.4 \\
                                                   & South Africa & 38 & 206 & 18.4 \\
\midrule
\textbf{Misinformation} & United States & 1560 & 7983 & 19.5 \\
                        & United Kingdom & 390 & 2114 & 18.4 \\
                        & India & 206 & 1255 & 16.4 \\
                        & Australia & 123 & 561 & 21.9 \\
                        & Israel & 41 & 266 & 15.4 \\
                        & South Africa & 39 & 206 & 18.9 \\
\midrule
\textbf{Privacy \& Security} & United States & 1151 & 7983 & 14.4 \\
                             & United Kingdom & 326 & 2114 & 15.4 \\
                             & India & 199 & 1255 & 15.9 \\
                             & Australia & 93 & 561 & 16.6 \\
                             & Israel & 50 & 266 & 18.8 \\
                             & South Africa & 31 & 206 & 15.0 \\
\midrule
\textbf{Human-Computer Interaction} & United States & 731 & 7983 & 9.2 \\
                                    & United Kingdom & 230 & 2114 & 10.9 \\
                                    & India & 103 & 1255 & 8.2 \\
                                    & Australia & 52 & 561 & 9.3 \\
                                    & Israel & 17 & 266 & 6.4 \\
                                    & South Africa & 17 & 206 & 8.3 \\
\bottomrule
\end{tabularx}

\caption{Distribution of articles reporting on each AI risk category by country, including the number of articles reporting on each risk category, total number of articles per country, and proportion of articles per country reporting on at least that risk category.}
\label{tab:risk_category_by_country_final}
\end{table}

\newpage

\twocolumn

\subsection{Prompt for filtering articles by content}\label{p1}

\begin{lstlisting}[language=Python, caption={Prompt to classify whether an article describes an impact of an AI system based on the conceptual definition provided in the prompt.},  numbers=none, breaklines=true]
### Context
{article_text}

### Definition
An impact refers to an effect, consequence, or outcome of an AI system (i.e., model or application) that positively or negatively affects individuals, organizations, communities, or society.

### Task
Based on the definition of an impact of an AI system provided to you, does the article above cover or describe at least one impact of an AI system? Answer Yes or No.

DO NOT explain yourself.
\end{lstlisting}

\subsection{Prompt for summarizing negative impacts from articles}\label{p2}
\begin{lstlisting}[language=Python, caption={Prompt to summarize the negative impacts of AI in an article based on the provided operational definition of a negative impact.},  numbers=none, breaklines=true]

Summarize ALL negative impacts that are ONLY relevant to the AI system or model described in the context provided to you. A negative impact refers to the set of risks or harms that have or may affect individuals, organizations, or communities in society, as a result of an AI system, or its use. Each summarized impacts must be 1 sentence long and must have sufficient details that are grounded in the context provided to you.

Format your response in a jsonl format {"negative_impacts": [list of impacts]}. If not impacts are present, output {"negative_impacts":[]}.

DO NOT make up details for impacts.
DO NOT interpret any details, share any highlights, or draw conclusions.
ONLY stick to the context provided to you.
DO NOT provide any other details in the answer besides the jsonl content.

### Context:
{article_text}
\end{lstlisting}

\subsection{Prompt for annotating negative impacts in corpus of articles}\label{p3}
\begin{lstlisting}[language=Python, caption={Prompt to scale up the annotation process of negative impacts of AI using GPT-4o as described in section \ref{4.2}.},  numbers=none, breaklines=true]

Task: Analyze the negative impact statements of AI technologies. A negative impact refers to the set of risks or harms that have or may affect individuals, organizations, or communities in society, as a result of an AI system, or its use. Each impact statement should be evaluated and categorized into ONLY one of the following 32 categories. For each impact statement listed, assign exactly one label that corresponds to its most appropriate classification in the same format as the original list. 
Categories :

1. ai_governance: negative impacts associated with governance policies and regulations related to the development, deployment, licensing, or moderation of AI technologies.

2. ai_incompetence: risks resulting from limitations and malfunctions of AI technologies that impact their performance.

3. authoritative_use_of_ai: risks associated with the potential misuse of artificial intelligence by governments in ways that may support authoritarian practices that violate human rights and civil liberties.

4. criminal_activities: risks associated with the misuse of AI technologies for online crimes such as cybercrimes or cyberattacks.

5. deception/manipulation: risks associated with the use of AI technologies for fraud, dishonest activities, sowing divisions, or misrepresenting individuals to influence or alter perceptions, behaviors, or mislead individuals or society.

6. discrimination/bias: risks of AI technologies generating outputs based on protected characteristics that result in unequal treatment or representation of individuals or social groups.

7. disruption_of_service: risks related to disruptions or reductions in the accessibility, availability, and functionality of AI systems.

8. economic_harm: risks posed by AI technologies to financial systems, labor market, and trading dynamics.

9. environmental: environmental and ecological risks arising from the energy consumption and resource intensive process required for the development, deployment, and operation of AI technologies.

10. ethical_impact: challenges related to inequalities in access to AI technologies, or in their development, deployment, and use, with a particular focus on issues of fairness, accountability, and transparency.

11. existential_threats: risks related to potential inequalities in accessing AI technologies, their development, deployment, and use, with a particular focus on issues of fairness, accountability, and transparency.

12. fundamental_rights: risks posed by AI technologies related to violating individual freedoms and rights, including freedom of expression and intellectual property.

13. information_risks: risks associated with AI hallucinations, including the generation of inaccurate information, low-quality AI-generated content, and fabrication of information by AI technologies.

14. hate/toxicity: risks associated with AI-generated content amplifying or spreading hateful, abusive, or offensive content.

15. humanness: risks related to the loss or diminishment of human qualities, such as creativity, emotional depth, and authentic interpersonal connections, due to AI technologies.

16. media_impacts: risks of AI technologies on the independence, integrity, and reliability of media and journalism.

17. mental_&_emotional: risks related to the psychological well-being and emotional health of individuals using or interacting with AI technologies.

18. operational_misuses: risks associated with the misuse of AI technologies in critical and highly regulated applications, such as unsafe autonomous operations, unreliable legal or military advice, or automated decision-making.

19. over-reliance: risks arising from over-relying on AI technologies in contexts that results in undermining human judgment, critical thinking, and decision-making.

20. political_useage: risks associated with the use of AI technologies to spread misinformation or disinformation, influence elections or politics, undermine democratic integrity, or disrupt social order.

21. privacy: risks related to unauthorized access, use, or disclosure of users' data and personal information.

22. safety_risks: risks of harming or endangering individuals' lives or safety arising from the malfunction or misuse of AI technologies.

23. security_risks: risks related to the threats and exploitation of vulnerabilities that compromise the confidentiality, integrity, or availability of AI technologies.

24. sexual_content: risks related to the non-consensual creation, distribution, or misuse of sexually explicit material or pornography using AI technologies.

25. structure/power: risks related to the concentration of power and AI resources or technologies among a few entities or governments, and its consequences on competition, collaboration, innovation, and safety of AI technologies.

26. technology_adoption: risks related to the adoption of AI technologies due to integration and usability challenges, or due to barriers faced by organizations and individuals in adopting AI technologies into their work.

27. user_experience: risks and issues that undermine the satisfaction satisfaction, trust, and interaction of the end-user with AI technologies.

28. violence_&_extremism: risks pertaining to the use of AI technologies or AI-generated content to incites violence, promotes extremist ideologies, or enables harmful activities such as weapon development or using AI in warfare.

29. defamation: risks involving reputational harms to individuals or organizations through AI-generated false or misleading statements, images, or representations.

30. child_harm: risks related to the misuse of AI technologies to harm or exploit children.

31. no_impact: refers to general statements that do not highlight potential or direct negative consequences, risks, or harms of AI systems.

32. other: Any risks or harms that do not fit into the above categories.

Output Format : Present the classified categories without any numbers and clean from whitespace. The categories should be selected from one of the above 33 categories.

Note: Ensure that each impact statement is classified under only one of the aforementioned 32 categories and that there is one
classification corresponding to each impact statement in the input. Do not generate any other text or include any additional details.
Do not make up categories.

Impact:
{summarized_impact}
\end{lstlisting}
\newpage

\end{document}